\documentclass[prl,floatfix,showpacs,twocolumn,preprintnumbers,amsmath,amssymb,superscriptaddress]{revtex4}
\usepackage[sort&compress]{natbib}
\usepackage{graphicx,float,datetime}
\usepackage{hyperref,wasysym}
\usepackage{epstopdf,subfigure}
\usepackage{enumitem}
\usepackage{bm}
\usepackage[utf8]{inputenc}

\newcommand\tauem{\overset{\textit{\tiny e-m}}{\mathcal{T}}}

\newcommand*\xbar[1]{%
  \hbox{%
    \vbox{%
      \hrule height 0.5pt 
      \kern0.5ex
      \hbox{%
        \kern-0.1em
        \ensuremath{#1}%
        \kern-0.1em
      }%
    }%
  }%
} 

\newcommand{\udt}[3]{#1^{#2}_{\phantom{#2}#3}}

\def\ber{\begin{eqnarray}}
\def\eer{\end{eqnarray}}
\def\beq{\begin{equation}}
\def\eeq{\end{equation}}

\newdateformat{mydate}{\THEDAY\hspace{3pt}\monthname[\THEMONTH] \THEYEAR}

\begin{document}

\begin{center}
\title{A Perturbative Approach to Neutron Stars in $f(T, \mathcal{T})-$Gravity}
\date{\mydate\today}
\author{Mark Pace\footnote{mark.pace.10@um.edu.mt}}
\affiliation{Department of Physics, University of Malta, Msida, MSD 2080, Malta}
\affiliation{Institute of Space Sciences and Astronomy, University of Malta, Msida, MSD 2080, Malta}
\author{Jackson Levi Said\footnote{jackson.said@um.edu.mt}}
\affiliation{Department of Physics, University of Malta, Msida, MSD 2080, Malta}
\affiliation{Institute of Space Sciences and Astronomy, University of Malta, Msida, MSD 2080, Malta}

\begin{abstract}
{
\noindent
We derive a Tolman-Oppenheimer-Volkoff equation in neutron star systems within the modified $f(T, \mathcal{T})$-gravity class of models using a perturbative approach. In our approach $f(T, \mathcal{T})$-gravity is considered to be a static spherically symmetric space-time. In this instance the metric is built from a more fundamental tetrad vierbein which can be used to relate inertial and global coordinates. A linear function $f = T(r) + \mathcal{T}(r) + \chi h(T, \mathcal{T}) + \mathcal{O}(\chi^{2})$ is taken as the Lagrangian density for the gravitational action. Finally we impose the polytropic equation of state of neutron star upon the derived equations in order to derive the mass profile and mass-central density relations of the neutron star in $f(T, \mathcal{T})$-gravity.
}
\end{abstract}

\pacs{04.40.Dg, 04.50.Kd}

\maketitle

\end{center}

\section{I. Introduction}\label{sec:intro}
\noindent
Recently it has been shown that the Universe is accelerating in its expansion \cite{garnavich1998supernova, riess1998observational}. The concept of the cosmological constant together with the inclusion of dark matter yield the $\Lambda$CDM model which explains a whole host of phenomena within the universe. \cite{feng2005dark, guo2005cosmological, boehmer2011existence}. We may also explain this acceleration by instead modifying the gravitational theory itself with alternative theories of gravity an example of which is $f(R)$-gravity \cite{de2010f, Capozziello:2011et, Cai:2015emx, Capozziello:2009nq}.
\\
\\
Our focus of this paper is on one alternative theory of gravity called $f(T)$-gravity which makes use of a ``teleparallel'' equivalent of GR (TEGR) \cite{iorio2012solar} approach, in which instead of the torsion-less Levi-Civita connection, the Weitzenb\"{o}ck connection is used, with the dynamical objects being four linearly independent vierbeins \cite{unzicker2005translation, hayashi1979new}. The Weitzenb\"{o}ck connection is curvature-free and describes the torsion of a manifold. 
\\
\\
The differences between $f(T)$ class of gravity and other gravity forms such as $f(R)$ and TEGR is in the choice of the function $f(T)$ which is taken \cite{Cai:2015emx}. Comparing $f(T)$-gravity with $f(R)$-gravity it is noted that $f(T)$-gravity cannot be reforumlated as a teleparallel action plus a scalar field through the conformal transformation due to the appearance of additional scalar-torsion coupling terms \cite{Ferraro:2011us, Ferraro:2011ks}. The obvious difference is that $f(T)$-gravity has a class of equations which is easier to work with because the field equations are second order rather than fourth-order like in $f(R)$-gravity class scenarios \cite{Cai:2015emx}. In $f(T)$-gravity more degrees of freedom are obtained which thus corresponds to one massive vector field \cite{Wang:2011xf, HamaniDaouda:2011iy}. 
\\
\\
We make use of a pure tetrad \cite{Tamanini:2012hg}, which means that the torsion tensor is formed by a multiple of the tetrad and its first derivative only. Under the assumption of invariance under general coordinate transformations, global Lorentz transformations, and the parity operation we construct the Lagrangian density from this torsion tensor \cite{Tamanini:2012hg, iorio2012solar, hayashi1979new, Capozziello:2009nq}. Also the Lagrangian density is second order in the torsion tensor \cite{iorio2012solar, hayashi1979new}. Thus $f(T)$-gravity generalises the above TEGR formalism, making the gravitational lagrangian a function of $T$ \cite{iorio2012solar, Cai:2015emx, Capozziello:2009nq}.
\\
\\
Our goal for this paper is to derive a working model for the TOV equations within a new modification of $f(T)$ class gravity, called $f(T, \mathcal{T})$-gravity in a perturbative manner. We make use of a perturbative approach due to the fact that a non physical assumption had to be taken whilst deriving the TOV equations in an analytical manner.
\\
\\
$f(T, \mathcal{T})$-gravity couples the gravitational sector and the standard matter one \cite{Cai:2015emx}. Instead of having the Ricci scalar coupled with the trace of the energy momentum tensor $\mathcal{T}$ as is done in $f(R, \mathcal{T})$-gravity, $f(T, \mathcal{T})$-gravity couples the torsion scalar $T$ with the trace of the matter energy-momentum tensor $\mathcal{T}$ \cite{Cai:2015emx, Planck:2013jfk, Capozziello:2009nq}. Recently a modification to this theory has been propose, that of allowing for a general functional dependence on the energy momentum trace scalar, $\udt{\mathcal{T}}{\mu}{\mu}=\mathcal{T}$ \cite{Cai:2015emx, Capozziello:2009nq}. 
\\
\\
Our interest is in studying the behaviour of spherically symmetric compact objects in this theory. We propose the use of a linear function, namely $f(T, \mathcal{T}) = \alpha T(r) + \beta \mathcal{T}(r) + \varphi$, where $\alpha$ and $\beta$ are arbitrary constants which may be varied to align our star's behaviour with current observations. $\varphi$ is then considered to be the cosmological constant. We consider the linear modification since it is the natural first functional form to consider, and the right place to start to understand how the trace of the stress-energy tensor might effect $f(T,\mathcal{T})$ gravity. In particular, our focus is on neutron stars in $f(T,\mathcal{T})$ gravity.
\\
\\
Besides the possibility of the existence of these exotic stars, this is also a good place to study the behavior of modified theories of gravity in terms of constraints. Moreover, this also opens the door to considerations of stiff matter in early phase transitions \cite{astashenok2015nonperturbative}.
\\
\\
The plan of this paper is as follows; in section 2 we go over the mathematical tools and give an overview of $f(T, \mathcal{T})-$gravity. In section 3 we discuss the rotated tetrad taken and discuss how the equations of motion in $f(T, \mathcal{T})-$gravity are derived perturbatively. In section 4 the two TOV equations are derived and discussed along with the schwarzschild solution, while the results are then used in section 5 where we output the numerical results given by the yielded TOV equations. Finally we discuss the results in section 6.

\section{II. $f(T, \mathcal{T})$-gravity Overview}\label{sec:overview}
\noindent
$f(T, \mathcal{T})$-gravity generalises $f(T)$-gravity and thus is based on the Weitzenbock's geometry. We will follow a similar notation style as that given in Ref. \cite{iorio2012solar, farrugia2016solar, paliathanasis2016cosmological, Capozziello:2011et, Cai:2015emx, Pace:2017aon}. Using: Greek indices $\mu, \nu, \dots$ and capital Latin indices $i, \kappa, \dots$ over all general coordinate and inertial coordinate labels respectively \cite{farrugia2016solar, paliathanasis2016cosmological, Capozziello:2011et, Cai:2015emx}.
\\
\\
\noindent
Torsion tensor \cite{farrugia2016solar, paliathanasis2016cosmological, Krssak:2015oua} is given by
\\
\begin{multline}
T^{\lambda}_{\enspace \mu \nu} \left( e^{\lambda}_{\enspace \mu}, \omega^{\lambda}_{\enspace i \mu} \right) = \partial_{\mu} e^{\lambda}_{\enspace \nu} - \partial_{\nu} e^{\lambda}_{\enspace \mu} + \\ \omega^{\lambda}_{\enspace i \mu} e^{i}_{\enspace \nu} - \omega^{i}_{\enspace \lambda \nu} e^{i}_{\mu},
\end{multline}

\noindent
where $\omega^{\lambda}_{\enspace i \mu}$ is the spin connection \cite{Krssak:2015oua}. The torsion tensor has vanishing curvature. Therefore by doing so all the information of the gravitational field is embedded in the torsion tensor \cite{Pace:2017aon}, while the gravitational Lagrangian is the torsion scalar \cite{Krssak:2015oua}. The contorsion tensor is then defined as

\begin{equation}
K^{\mu \nu}_{\enspace \enspace \rho} = - \dfrac{1}{2} \left(T^{\mu \nu}_{\enspace \enspace \rho} - T^{\nu \mu}_{\enspace \enspace \rho} - T_{\rho}^{\enspace \mu \nu} \right),
\end{equation}
\noindent
while the superpotential of teleparallel gravity is defined by \cite{farrugia2016solar, paliathanasis2016cosmological}

\begin{equation}
S_{\rho}^{\enspace \mu \nu} = \dfrac{1}{2} \left( K^{\mu \nu}_{\enspace \enspace \rho} + \delta^{\mu}_{\rho} T^{\alpha \nu}_{\enspace \enspace \alpha} - \delta^{\nu}_{\rho} T^{\alpha \mu}_{\enspace \enspace \alpha} \right).
\end{equation}
\noindent
Unlike the contorsion tensor, the superpotential tensor does not have any apparent physical meaning, instead is it purely introduced to reduce the size of the Lagrangian.
\\
\\
The torsion scalar \cite{farrugia2016solar, paliathanasis2016cosmological, Pace:2017aon} is then given as

\begin{equation} \label{Torsion_scalar_def}
T = S_{\rho}^{\enspace \mu \nu} T^{\rho}_{\enspace \mu \nu}.
\end{equation} 

\noindent
As in the analogous $f(R,T)$ theories \cite{harko2011f}, we further generalised upon the gravitational lagrangian by taking an arbitrary function $f$ and thus giving \cite{nassur2015early, harko2014f}

\begin{equation}
\label{action}
S = - \dfrac{1}{16 \pi G} \int d^{4} x e \left[ f(T, \mathcal{T}) + \mathcal{L}_{m}  \right].
\end{equation}
The function $f(T, \mathcal{T})$ is taken to be equal to $T(r) + \mathcal{T}(r) + \chi h(T, \mathcal{T}) + \mathcal{O}(\chi^{2})$ where $\chi$ is a very small paramter which will aid in differentiating between zeroth and first order term \cite{arapouglu2011constraints}, and $h(T, \mathcal{T})$ is an arbitatary function of the torsion scalar $T$ and the trace $\mathcal{T}$ of the energy momentum tensor $\tauem$ given by $\mathcal{T} = \delta^{\nu}_{\mu}\mathcal{T}_{\nu}^{\;\mu}$. $\mathcal{L}_m$ is the matter Lagrangian density \cite{nassur2015early, Pace:2017aon}. In this instance $f$ is an arbitrary function of the torsion scalar $T$ and the trace of the energy-momentum tensor $\mathcal{T}$ \cite{nassur2015early}. The variation of the action defined in Eq.(\ref{action}) with respect to the tetrad leads to the field equations \cite{arapouglu2011constraints}

\begin{widetext}
\begin{multline}
\label{fieldequations}
e^{\rho}_{i} S_{\rho}^{\; \mu \nu} \partial_{\mu} T \chi h_{TT} + e^{\rho}_{i} S_{\rho}^{\; \mu \nu} \chi h_{T \mathcal{T}} \mathcal{T} + e^{-1} \partial_{\mu} \left( e e^{\rho}_{i} S_{\rho}^{\; \mu \nu} \right) (1+ \chi h_{T})  + e^{\mu}_{i} T^{\lambda}_{\; \mu \kappa} S_{\lambda}^{\nu \kappa} (1+ \chi h_{T}) \\ - \dfrac{e^{\nu}_{i} T(r) + \mathcal{T}(r) + \chi h(T, \mathcal{T}) }{4} + (1 + \chi h_{T}) \omega^{i}_{\enspace \lambda \nu} S_{i}^{\enspace \nu \mu} - \dfrac{(1+ \chi h_{\mathcal{T}})}{2} \left( e^{\lambda}_{i}  \mathcal{T}_{\lambda}^{\;\nu} + p(r) e^{\nu}_{i} \right) = - 4 \pi e^{\lambda}_{i} \tauem^{\; \nu}_{\lambda},
\end{multline}
\end{widetext}
\noindent
where $h_{T} = \dfrac{\partial h}{\partial T}$, $h_{\mathcal{T}} = \dfrac{\partial h}{\partial \mathcal{T}}$, and $ h_{T \mathcal{T}} = \dfrac{\partial^2 h}{\partial T \partial \mathcal{T}}.$


\section{III. Perturbative Equations of Motion in $f(T, \mathcal{T})$-gravity}\label{sec:overview}
\noindent
In perturbative theory the field equations may be expanded perturbatively in $\chi$ \cite{arapouglu2011constraints} and therefore the metric components take on the expansions $A(r)_{\chi} = A(r) + \chi A(r)_{1} + \dots$ and $B(r)_{\chi} = B(r) + \chi B(r)_{1} + \dots$ \cite{arapouglu2011constraints}. The energy-momentum tensor in the field equations, is still the energy-momentum tensor of the perfect fluid. The hydrodynamic quantities are also defined perturbatively by $\rho(r)_{\chi} = \rho(r) + \chi \rho(r)_{1} + \dots$ and $p(r)_{\chi} = p(r) + \chi p(r)_{1} + \dots$ \cite{arapouglu2011constraints}.
\\
\\
A spherically symmetric metric which has a diagonal structure is considered for our system \cite{deliduman2011absence},

\begin{equation} \label{metric}
ds^{2} = - e^{A(r)_{\chi}} dt^{2} + e^{B(r)_{\chi}} dr^{2} + r^{2} d\theta^{2} + r^{2} \sin^{2} d \phi^{2},
\end{equation}

\noindent
and we consider the fluid inside the star to be that of a perfect fluid which yields a diagonal energy-momentum tensor

\begin{equation}
\tauem^{\; \nu}_{\lambda} = diag ( - \rho(r)_{\chi}, p(r)_{\chi}, p(r)_{\chi}, p(r)_{\chi}),
\end{equation}

\noindent
where $\rho(r)_{\chi}$ and $p(r)_{\chi}$ are the energy density and pressure of the fluid respectivel \cite{deliduman2011absence}. These also make up the matter functions which, along with the metric functions, $A(r)$ and $B(r)$, are also taken to be independent of time \cite{Pace:2017aon}. Thus the system is taken to be in equilibrium \cite{boehmer2011existence, deliduman2011absence}. 
\\
\\
The equation of conservation of energy is given by

\begin{equation} \label{conservation}
\dfrac{dp(r)}{dr} = - (\rho(r) + p(r)) \dfrac{dA(r)}{dr}.
\end{equation}

\noindent
Following Ref. \cite{Tamanini:2012hg} the following rotated tetrad is used

\begin{widetext}
\[
e_{\mu}^{a}   =
 \left( \begin{array}{cccc}
e^{\dfrac{A(r)_{\chi}}{2}} & 0 &  0 & 0 \\
0 & e^{\dfrac{B(r)_{\chi}}{2}} \sin{\theta} \cos{\phi} & e^{\dfrac{B(r)_{\chi}}{2}} \sin{\theta} \sin{\phi} & e^{\dfrac{B(r)_{\chi}}{2}} \cos{\theta} \\
0 & -r \cos{\theta} \cos{\phi} & -r \cos{\theta} \sin{\phi} & r \sin{\theta}\\
0 & r \sin{\theta} \sin{\phi} & - r \sin{\theta} \cos{\phi} & 0 \end{array} \right)\]
\end{widetext}

\noindent
This form of vierbein is considered because it allows us more degrees of freedom \cite{Faraoni:2000wk} and it allows us to acquire a static and spherically symmetric wormhole solution in our standard formulation of $f(T, \mathcal{T})$-gravity \cite{Faraoni:2000wk, Paliathanasis:2014iva}.
\\
\\
Also because this is a pure form of tetrad \cite{Krssak:2015oua}, the spin connection elements of the tetrad vanish and thus ensure that the spin connection terms need not be included \cite{Krssak:2015oua}.
\\
\\
Inserting this vierbein into the field equations, from Eq.(\ref{Torsion_scalar_def}) we get the resulting torsion scalar

\begin{multline} \label{torsionscalar}
T(r) = \\ \dfrac{2 e^{-B(r)}}{r^2} \left(1-e^{\dfrac{B(r)}{2}} \right) \left(1 - e^{\dfrac{B(r)}{2}} + r A'(r) \right),
\end{multline}

\noindent
where the prime denotes derivative with respect to $r$. The resulting field equation components turn out to be.
\\
\\
\begin{widetext}
The $t - t$ component, given by $i = \nu = 0$ results in

\begin{multline} \label{f_TT_perturbative_00}
4 \pi \rho(r)_{\chi} = \dfrac{T(r)_{\chi} + \mathcal{T}(r)_{\chi} + \chi h}{4} + \dfrac{e^{-B(r)}}{2 r^{2}} \left(1 + \chi h_{T} \right) \left[ -2 + 2e^{\dfrac{B(r)}{2}} + r A'(r) \left(e^{\dfrac{B(r)}{2}} - 1 \right) + r B'(r) \right] \\
+ \dfrac{\left( 1 + \chi h_{\mathcal{T}} \right)}{2} \left( \rho(r) - p(r) \right) + \dfrac{e^{-B(r)} \chi}{r} \left(e^{\dfrac{B(r)}{2}} - 1 \right) \left( h_{TT} T'(r) + h_{T \mathcal{T}} \mathcal{T}'(r) \right).
\end{multline}

While the $r - r$ component, given by $i = \nu = 1$ results in

\begin{multline} \label{f_TT_perturbative_11}
4 \pi p(r)_{\chi} = \dfrac{T(r)_{\chi} + \mathcal{T}(r)_{\chi} + \chi h}{4} \\
+ \dfrac{e^{-B(r)}}{2 r^{2}} \left(1 + \chi h_{T} \right) \left[ 2 \left(e^{\dfrac{B(r)}{2}} - 1 \right) + r A'(r) \left(e^{\dfrac{B(r)}{2}} - 2 \right) \right] - p(r) \left(1 + \chi h_{\mathcal{T}} \right).
\end{multline}
\end{widetext}
Note that the zeroth order quantities are given without a subscript.

\section{IV. Perturbative Derivation in $f(T, \mathcal{T})-$Gravity}\label{sec:derivation}
\noindent
We will now make use of the equations of motion given by Eq. (\ref{f_TT_perturbative_00}) and Eq. (\ref{f_TT_perturbative_11}) by first considering a solution for $\rho(r)_{\chi}$ and $p(r)_{\chi}$ up to order $\chi$. The zeroth order quantities are considered from these two equations and given by

\begin{multline}
4 \pi \rho(r) = \dfrac{T(r) + \mathcal{T}(r)}{4}  + \dfrac{e^{-B(r)}}{2 r^{2}} \bigg[ -2 + 2e^{\dfrac{B(r)}{2}} \\ + r A'(r) \left(e^{\dfrac{B(r)}{2}} - 1 \right) + r B'(r) \bigg]  + \dfrac{\left( \rho(r) - p(r) \right)}{2},
\end{multline}

and

\begin{multline} 
4 \pi p(r) = \dfrac{T(r) + \mathcal{T}(r)}{4}
+ \dfrac{e^{-B(r)}}{2 r^{2}} \bigg[ 2 \left(e^{\dfrac{B(r)}{2}} - 1 \right)\\ + r A'(r) \left(e^{\dfrac{B(r)}{2}} - 2 \right) \bigg] - p(r).
\end{multline}

At this point the torsion scalar given by Eq. (\ref{torsionscalar}) and $\mathcal{T}(r) = \rho(r) - 3p(r)$ are inserted into the two equations which after manipulation result in

\begin{multline} \label{f_TT_perturbative_00_zeroth}
4 \pi \rho(r) = \dfrac{e^{-B(r)}}{2 r^{2}} \left( -1 + e^{B(r)} + r B'(r) \right) \\ + \dfrac{1}{4} \left(3 \rho(r) - 5p(r) \right),
\end{multline}

and

\begin{multline} \label{f_TT_perturbative_11_zeroth}
4 \pi p(r) = \dfrac{e^{-B(r)}}{2 r^{2}} \left( -1 + e^{B(r)} - r A'(r) \right) \\ + \dfrac{1}{4} \left(\rho(r) - 7p(r) \right),
\end{multline}

respectively.
\\
\\
At this point it is convenient to take a mass parameter ansatz. The solution is assumed to have the same form of the exterior solution for the metric function $B_{\chi}$. In order to render a metric ansatz in line with the Schwarzschild metric we take the following \cite{arapouglu2011constraints}.

\begin{equation} \label{schwarzschild_solution_perturbative}
e^{-B(r)_{\chi}} = 1 - \dfrac{\Omega M(r)_{\chi}}{r} + \epsilon(r)_{\chi},
\end{equation}
\noindent
where $\Omega$ is an arbitrary constant and $\epsilon(r)$ is taken to be a function of $r$. Similar to $\rho_{\chi}$, $M(r)_{\chi}$ is expanded in $\chi$ as $M_{\chi} = M + \chi M_{1} + \dots$, \cite{arapouglu2011constraints} where $M$ is the zeroth order solution.
\\
\\
Taking a derivative of $M_{\chi}$ with respect to $r$ the following is obtained

\begin{multline} \label{dm_dr_perturbative}
\dfrac{d M_{\chi}}{dr} = \dfrac{1}{\Omega} \bigg(1 - e^{-B(r)_{\chi}} + \epsilon(r)_{\chi} \\ + e^{-B(r)_{\chi}}r B'(r)_{\chi} + r \epsilon'(r)_{\chi} \bigg).
\end{multline}

\noindent
Now we focus on the first equation of motion given by Eq. (\ref{f_TT_perturbative_00}) where we insert the torsion scalar equation and the energy momentum components and thus obtain the following equation

\begin{widetext}
\begin{multline}
4 \pi \rho(r)_{\chi} = \dfrac{\Omega}{2 r^{2}} \dfrac{dM_{\chi}}{dr} - \dfrac{(\epsilon_{\chi}(r))'}{2r^{2}} + \dfrac{1}{4} \left( 3 \rho(r)_{\chi} - 5 p(r)_{\chi} \right) \\ + 
\chi \Bigg\{ \dfrac{h}{4} + \dfrac{e^{-B(r)}h_{T}}{2 r^{2}} \left[ -2 + 2 e^{\dfrac{B(r)}{2}} + r A'(r) \left( e^{\dfrac{B(r)}{2}} - 1 \right) + r B'(r) \right] \\ + \dfrac{h_{\mathcal{T}}}{2} \left(\rho(r) - p(r) \right) + \dfrac{e^{-B(r)}}{r} \left( e^{\dfrac{B(r)}{2}} - 1 \right) \left( h_{TT} T'(r) + h_{T \mathcal{T}} \mathcal{T}'(r) \right) \Bigg\}.
\end{multline}
\end{widetext}

\noindent
Here we invoke a linear parameter for $h$, given by $\alpha T(r) + \beta \mathcal{T}(r) + \varphi$ which after being inserted into this equation and further reduced yields

\begin{widetext}
\begin{multline}
\dfrac{dM_{\chi}}{dr} = \dfrac{8 \pi r^{2} \rho(r)_{\chi}}{\Omega} + \dfrac{(\epsilon_{\chi}(r))'}{2r^{2}} - \dfrac{r^{2}}{2 \Omega} \left( 3 \rho(r)_{\chi} - 5 p(r)_{\chi} \right) \\ - \dfrac{r^{2} \chi}{2 \Omega} \Bigg\{ \dfrac{\alpha e^{-B(r)}}{r^{2}} \left[ 2 e^{\dfrac{B(r)}{2}} - 3 + e^{B(r)} + r A'(r) \left( e^{\dfrac{B(r)}{2}} - 1 \right) + 2 r B'(r) \right] \\ + \varphi + \beta \left( 3 \rho(r) - 5 p(r) \right)   \Bigg\}.
\end{multline}
\end{widetext}

\noindent
The main task at this point is to reduce the values of $A'(r)$ and $B'(r)$ where the definitions given by Eq. (\ref{f_TT_perturbative_00_zeroth}), Eq. (\ref{f_TT_perturbative_11_zeroth}), and Eq. (\ref{schwarzschild_solution_perturbative}) will be substituted and thus resulting in our first TOV equation

\begin{widetext}
\begin{multline} \label{Perturbative_dmdr}
\dfrac{dM_{\chi}}{dr} = \dfrac{8 \pi r^{2} \rho(r)_{\chi}}{\Omega} + \dfrac{(\epsilon_{\chi}(r))'}{2r^{2}} - \dfrac{r^{2}}{2 \Omega} \left( 3 \rho(r)_{\chi} - 5 p(r)_{\chi} \right) - \dfrac{r^2 \chi \varphi}{2 \Omega} \\ 
+ \dfrac{\chi \alpha}{4 \Omega} \left( 1 - \dfrac{\Omega M(r)}{r} + \epsilon(r) \right)^{\dfrac{1}{2}} \Bigg\{ - 2 + r \bigg[ r + r \epsilon(r) - M(r) \Omega \bigg]^{-1} \bigg[ r^2 \left(p(r) (7+16\pi) - \rho(r) \right)-2 \bigg] \\
+ \left( 1 + \epsilon(r) - \dfrac{\Omega M(r)}{r} \right)^{- \dfrac{1}{2}} \bigg[ 4 - r^2 \left(p(r) (17+16\pi) + \rho(r) (32 \pi - 7) \right) \bigg] \Bigg\}.
\end{multline}
\end{widetext}

Now we shift our focus into deriving the pressure-radius relation of the TOV equations. For this purpose Eq. (\ref{f_TT_perturbative_11}) is considered where a similar treatment will be given i.e. we substitute the torsion scalar equation and the energy momentum definition to give

\begin{widetext}
\begin{multline}
A'(r)_{\chi} = e^{B(r)_{\chi}} \Bigg\{ \dfrac{1}{r} \left( 1 - e^{-B(r)_{\chi}} \right) + \dfrac{r}{2} \left( \rho(r)_{\chi} - 7 p(r)_{\chi} \right) - 8 \pi p(r)_{\chi} r
\\
+ 2 r \chi \bigg\{ \dfrac{h}{4} + \dfrac{h_{T} e^{-B(r)_{\chi}}}{2 r^{2}} \bigg[ 2 \left( e^{\dfrac{B(r)}{2}} - 1 \right) + r A'(r) \left( e^{\dfrac{B(r)}{2}} - 2 \right) - h_{\mathcal{T}} p(r) \bigg\} \Bigg\}.
\end{multline}
\end{widetext}
\noindent
Inserting the definition of $h$ and Eq. (\ref{schwarzschild_solution_perturbative}) and reducing further yields

\begin{multline}
A'(r)_{\chi} = \Bigg\{ 8 p(r) \pi r \left( \alpha \chi - 1 \right) + \dfrac{\Omega M(r)}{r^2} - \dfrac{\epsilon(r)}{r} + \dfrac{r \chi \varphi}{2} \\
- \dfrac{r}{2} \left( 7p(r) - \rho(r) \right) \left( 1 + \beta - \alpha \chi \right)  \Bigg\} \bigg[ 1 - \dfrac{\Omega M(r)_{\chi}}{r} + \epsilon(r)_{\chi} \bigg]^{-1}.
\end{multline}

This result is then inserted into the continuity equation given by Eq. (\ref{conservation}) and thus results in the second TOV equation required

\begin{widetext}
\begin{multline} \label{Perturbative_dpdr}
\dfrac{dp(r)_{\chi}}{dr}  = - \left( \rho(r)_{\chi} + p(r)_{\chi} \right) \Bigg\{ 8 p(r) \pi r \left( \alpha \chi - 1 \right) + \dfrac{\Omega M(r)}{r^2} - \dfrac{\epsilon(r)}{r} + \dfrac{r \chi \varphi}{2} \\
- \dfrac{r}{2} \left( 7p(r) - \rho(r) \right) \left( 1 + \beta - \alpha \chi \right)  \Bigg\} \bigg[ 1 - \dfrac{\Omega M(r)_{\chi}}{r} + \epsilon(r)_{\chi} \bigg]^{-1} .
\end{multline}
\end{widetext}

\section{V. Numerical Modeling of Neutron Stars}\label{sec:modelling}
\noindent
Through Eq. (\ref{Perturbative_dmdr}) and Eq. (\ref{Perturbative_dpdr}) any spherically symmetric mass in $f(T,\mathcal{T})$-gravity can be investigated in terms of its physical properties. In order to obtain a mass profile relation for the TOV equations, we numerically integrate our TOV equations of stellar structure to build models of neutron stars in $f(T,\mathcal{T})$-gravity. Here we take the relativistic energy density $\rho(r)$ as equal to

\begin{equation}
\rho(r) = \rho_{0}(r) + \dfrac{p(r)}{\Gamma-1},
\end{equation}
\noindent
where $\rho_{0}(r)$ is the rest matter density \cite{Raithel:2016bux}. We take the initial conditions as equal to $m(0)=0$ and $p(0)=K \rho_{0,c}^{\Gamma}$ where $\rho_{0,c}^{\Gamma}(r)=10^{15} \text{gm/cm}^3$ is the central density \cite{Raithel:2016bux}. We take $\Gamma = 4/3$ and $G = 1$ and $c = 1$ \cite{Raithel:2016bux}.
\\
\\
The value of $\Omega$ is taken to be $2$ as the literature in Ref. \cite{arapouglu2011constraints} suggests and the value of $\varphi$ is taken to be the cosmological constant, as $2.036 \times 10^{-35}$ \cite{carmeli2001value}. Here we also consider our value of $\chi$ as being a very small but non-zero value $\sim 10^{-12} cm^{-2}$ \cite{arapouglu2011constraints}. The value of $\alpha$ in Eq. (\ref{Perturbative_dmdr}) and Eq. (\ref{Perturbative_dpdr}) is taken to be $-1$ and the value of $\beta$ is varied. We vary the value of $\beta$ so as to manipulate the dominance of the function $\mathcal{T}(r)$. 

\subsection{A. Mass Profile Curve}
\begin{center}
\begin{figure}[h!] 
\includegraphics[scale=0.3]{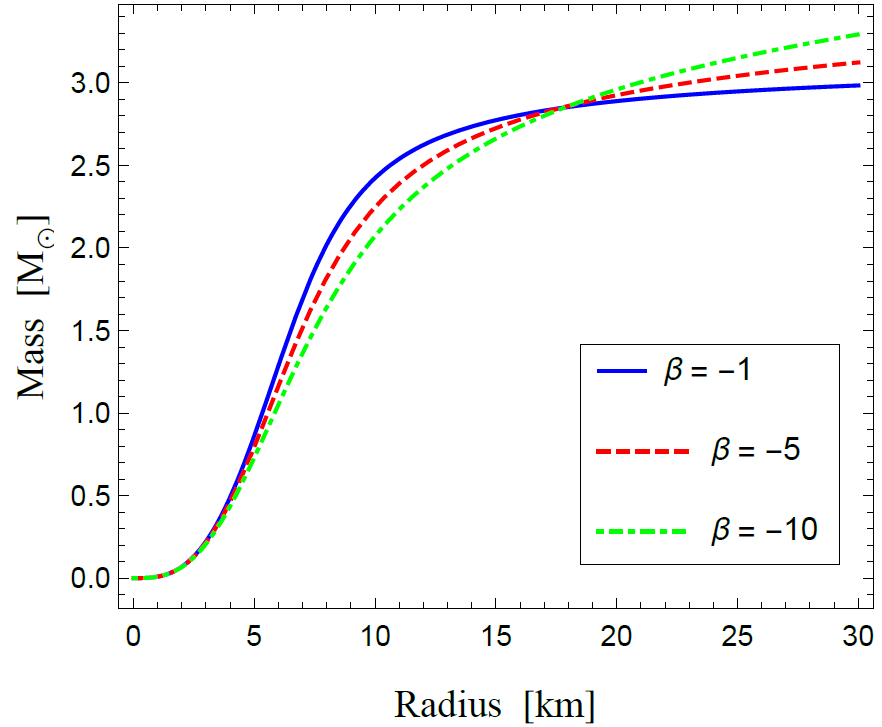}
\caption{Mass profile graph of a neutron star where $\alpha=-1$ and three different variations of $\beta$. The value of $\Gamma = 4/3$, $\Omega=2$ \cite{arapouglu2011constraints} and $\varphi = 2.036 \times 10^{-35}$ \cite{carmeli2001value}} \label{figMR}
\vspace{-0.5cm}
\end{figure}
\end{center}

\noindent
In Fig.(\ref{figMR}) we show the mass profile curve of a neutron star. We take $\beta = -1$ to include the GR case at this order of the perturbation and contrast with decreasing values of $\beta$.
\\
\\
As the function $\mathcal{T}(r) = \rho(r) - 3p(r)$ \cite{harko2014f} is included i.e. when $\beta = -5$ is taken, a similar mass profile is generated from the TOV equations. Fig.(\ref{figMR}) shows that the neutron star at first appears to be smaller in nature, however around the $17 km$ mark it surpasses the neutron star generated by the GR case to yield a larger stellar structure.
\\
\\
Current observations shown in the literature by \cite{Antoniadis:2013pzd} state that such massive neutron stars may exist. In fact the figures show that there is $\approx 4.67 \%$ increase in the maximum mass value of the neutron star gained.
\\
\\
We further magnify this effect by taking $\beta = -10$ and as may be seen from the Fig.(\ref{figMR}), the neutron star at first behaves exactly like the previous case and again surpasses the other cases at the $17 km$ mark to yield a more massive neutron star. In fact by considering $\beta = -10$ the allowable maximum mass of the neutron star is increased by $5.43 \%$ over the $\beta = -5$ case and $10.35 \%$ over the GR case.
\\
\\
In order to better understand the physical behaviour of neutron stars we investigate how mass varies over radius for different settings of $\beta$ in Fig.(\ref{figMR}). Whichever values of $\beta$ are taken the graphical output is similar in structure to that of GR. Another aspect to note is that the general behavior of the stellar system remains the same however a new degree of freedom is allowed depending on the maximum mass and radius of these stars.
\\
\\
In contrast the results gained in this paper are similar to those gained by various other authors such as found in Ref. \cite{tooper1964general, komatsu1989rapidly, stergioulas2003rotating, chandrasekhar1964dynamical}. The resulting mass profile curves behave in a similar manner where mass steadily increases with radius to plateau at an instance.

\subsection{B. Radius-Central Density Curve}

\begin{center}
\begin{figure}[h!] 
\includegraphics[scale=0.325]{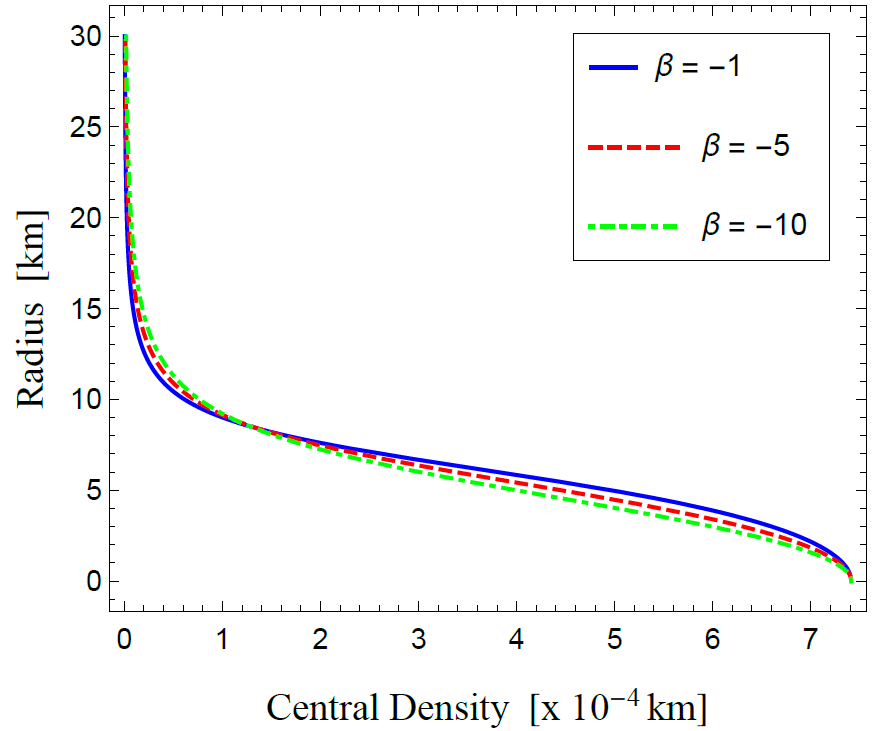}
\caption{Radius-central density graph of a neutron star where $\alpha=-1$ and three different variations of $\beta$. The value of $\Gamma = 4/3$, $\Omega=2$ \cite{arapouglu2011constraints} and $\varphi = 2.036 \times 10^{-35}$ \cite{carmeli2001value}} \label{figDR}
\vspace{-0.5cm}
\end{figure}
\end{center}

\noindent
Fig.(\ref{figMR}) is heavily dependent on the central density, not in behaviour but in terms of the particular values being produced. To contrast this we plot the radius-central density curve which was generated from the TOV equations. This plot is given by Fig.(\ref{figDR}).
\\
\\
Again we contrast with the GR case when taking $\beta = -1$. When we decrease the value of $\beta$ by taking $\beta = -5$ we may note that the central density figure is significantly lower than that of the GR case. It is noted that in both cases the curves decrease at the same rate with radius and curve at the same instance to intersect at a point close to $1 km$.
\\
\\
To magnify this we again take a lower value of $\beta$ by taking $\beta = -10$, this shows that the central density declines somewhat rapidly at first initially with radius and then slowly declines steadily similar to the previous cases however it reaches the lowest value the fastest out of the three cases considered. Thus this shows that a slightly larger neutron star is allowed in such a gravitational framework.

\noindent
Much like the mass profile curves the results gained from Ref. \cite{tooper1964general, komatsu1989rapidly, stergioulas2003rotating, chandrasekhar1964dynamical} exhibit similar results as gained in this study. The central density value increases significantly with radius at first to gain a maximum value as we get closer to the centre of the star.

\section{VI. Conclusion}\label{sec:conc}
\noindent
In this study the TOV equations are derived in a perturbative way for $f(T, \mathcal{T})-$gravity. Later the two equations are applied to a polytropic equation of state which yielded the characteristics of the neutron star in such a gravitational framework.
\\
\\
Our main goal throughout this research is to derive a working model which involved little to no assumptions in the derivation. We also wanted to retain and include as many general terms as possible. We did this also because we would like to further fine tune our results to current observations.
\\
\\
A reasonable boundary condition was taken in order to solve the TOV equations by numerical techniques. We apply the polytropic equation of state in order to reduce our TOV equations from a four variable equation to a three variable equation by making one of the variables dependent on the others.
\\
\\
Our approach considered a value of $\chi$ which is non-zero however very small $\sim 10^{-12} cm^{-2}$. The literature shown in Ref. \cite{arapouglu2011constraints} shows that the typical value of the Ricci curvature is calculated to be roughly on a similar order. Thus assuming our value of $\chi$ to be so small is reasonable.
\\
\\
Our graphical representations are inspired by the work carried out by the authors of Ref. \cite{Kpadonou:2015eza, Cai:2015emx}. The graphs show that a larger neutron star is allowed in such a gravitational framework. We vary the values of $\beta$ accordingly to output the variations occurring when we include the $\mathcal{T}(r)$ term. By taking a lower value of this term we note that it allows for a larger neutron star. This value will require future fine tuning in order to align with current observations.
\\
\\
More values of $\beta$ were considered in testing. The yielded results showed that when positive values of $\beta$ were considered no tangible neutron star would be yielded in such a gravitational framework. When lower values of $\beta$ were considered the yielded stellar structures did not behave in accordance to the theory as explained in Ref. \cite{tooper1965adiabatic}. Thus the range of values for $\beta$ considered to yield a tangible and proper neutron star would be $-10 \ll \beta \ll -1$.
\\
\\
There has not yet been many extensive studies conducted where the linear Lagrangian approach is considered as is done in this manuscript. However there have been cosmological studies that reconstruct the Lagrangian for various state parameter conditions \cite{harko2014f}.
\\
\\
For future work we also hope to derive the TOV equations using a non linear Lagrangian, however till now we have not been able to yield working TOV equations. We also hope to derive the TOV equations in $f(T, \mathcal{T})-$gravity in an analytical manner.

\section{VII. Acknowledgments}\label{sec:Acknow}
\noindent
\textit{The research work disclosed in this publication is funded by the ENDEAVOUR Scholarship Scheme (Malta). The scholarship may be part-financed by the European Union - European Social Fund (ESF) under Operational Programme II - Cohesion Policy 2014-2020, "Investing in human capital to create more opportunities and promote the well being of society"}.



\end{document}